# On the perceived relevance of critical internal quality attributes when evolving software features

Eduardo Fernandes
*Department of Computer Science*
*Federal University of Minas Gerais (UFMG)*
Belo Horizonte, Brazil
https://orcid.org/0000-0002-1222-2247

Marcos Kalinowski
*Informatics Department*
*Pontifical Catholic University of Rio de Janeiro (PUC-Rio)*
Rio de Janeiro, Brazil
https://orcid.org/0000-0003-1445-3425

*Abstract*—Several refactorings performed while evolving software features aim to improve internal quality attributes like cohesion and complexity. Indeed, internal attributes can become critical if their measurements assume anomalous values. Yet, current knowledge is scarce on how developers perceive the relevance of critical internal attributes while evolving features. This qualitative study investigates the developers' perception of the relevance of critical internal attributes when evolving features. We target six class-level critical attributes: low cohesion, high complexity, high coupling, large hierarchy depth, large hierarchy breadth, and large size. We performed two industrial case studies based on online focus group sessions. Developers discussed how much (and why) critical attributes are relevant when adding or enhancing features. We assessed the relevance of critical attributes individually and relatively, the reasons behind the relevance of each critical attribute, and the interrelations of critical attributes. Low cohesion and high complexity were perceived as very relevant because they often make evolving features hard while tracking failures and adding features. The other critical attributes were perceived as less relevant when reusing code or adopting design patterns. An example of perceived interrelation is high complexity leading to high coupling.

## I. INTRODUCTION

Several techniques have been proposed to monitor symptoms of source code degradation [1], [2], [3], [4]. Many of these techniques rely on measuring the code structure aimed at spotting degraded code structures and design [1]. Using internal quality attributes is a major technique. Each internal attribute captures a particular aspect of internal software quality [5]. Two examples of internal attributes are cohesion [6] and complexity [7]. While cohesion captures the interrelation degree of attributes and methods within a class [6], [2], complexity captures the cognitive difficulty of understanding code elements [7], [2].

Assessing metric values may assist in managing critical internal attributes, e.g., low cohesion and high complexity. A critical attribute is an internal attribute whose metrics used for capturing it assume anomalous values in comparison to the reference value [5], [8]. This paper targets metrics that become critical as their values increase, e.g., Lack of Cohesion (LCOM2) [6] and Cyclomatic Complexity (CC) [7] whose high values suggest classes with non-cohesive features or a high complexity [6], [7]. Critical attributes are symptoms that developers should consider managing – either mitigating or fully addressing – for the sake of software evolution [5].

Refactorings are largely advertised as effective means to help manage degradation symptoms [9], [10], including critical attributes. However, until recently, there was little empirical evidence on how refactorings affect internal attributes. A recent study addressed this literature gap in a large quantitative study [5] targeting five internal attributes: cohesion, complexity, coupling, inheritance, and size. That study suggests the refactoring effect on improving, worsening, or keeping internal attributes unaffected is diverse. Popular refactoring types like Extract Method and Move Method [9], [10] improve one or another attribute while worsening others. Developers should carefully apply these refactorings to avoid the unexpected worsening of internal attributes, thereby potentially harming software evolution.

Due to the quantitative nature of the previous study mentioned above [5], an important question remained unanswered: *how much (and why) are critical attributes perceived as relevant by developers while evolving features?* Answering this question is essential to assist refactorings along with software evolution. This knowledge could help in managing critical attributes, while preventing the worsening of potentially relevant attributes.

This paper presents an industrial case study, based on case study guidelines [11], that addresses the aforementioned gap. We investigate the relevance degree reported by developers on critical attributes spotted by the five internal attributes assessed by the previous quantitative study [5]: cohesion, complexity, coupling, inheritance, and size. We elicit reasons why developers find each critical attribute relevant (or irrelevant) while evolving features. We also reveal some interrelations of critical attributes that may be relevant while evolving features.

We recruited two development teams of the ExACTa PUC-Rio industry-academia Research and Development (R&D) collaboration initiative working on projects with Petrobras. Each team engaged in one focus group session [12]. Developers discussed the relevance of six critical attributes: low class cohesion, high class complexity, high class coupling,

Funded by CNPq (grant 312827/2020-2) and FAPERJ (grant 200.773/2019).

large class hierarchy depth, large class hierarchy breadth, and large class size. We refer to *relevance* as the need for either mitigating or fully addressing critical attributes while evolving features.

Our study results suggest the following. First, low class cohesion and high class complexity are perceived as relevant while evolving features. This result stands out because popular refactorings [9], [10], such as Extract Method and Move Method, often worsen cohesion [5]. Second, high class coupling, large class hierarchy depth, large class hierarchy breadth, and large size are not necessarily perceived as relevant by developers while evolving features. This result is curious because refactorings rarely worsen inheritance, and only a few refactoring types worsen coupling and size (e.g., Pull Up Method [5]). Third, we found relationships between certain critical attributes, e.g., high class complexity may lead to high class coupling. These results could support the design of refactoring tools that, for enhancing code structures, optimize certain critical attributes to the detriment of others based on their perceived relevance for developers.

**Data availability statement:** We made our study artifacts available online [13].

## II. BACKGROUND

### A. Critical attributes

Anomalous metric values can help monitor internal quality attributes [14], [15], [5]. Each internal attribute concerns an internal property of the system. Table I defines the five internal attributes we investigate as defined in a recent work [5]. These definitions are limited to the scope of our work. E.g., complexity [7] could target other granularities rather than class, but we are concerned about the class complexity.

Table I
METRICS GROUPED BY INTERNAL ATTRIBUTE

| Attribute | Definition | Metric Name |
|---|---|---|
| Cohesion | *Cohesion* [6], [2] captures the interrelation degree of code elements, i.e. attributes and methods, that constitute a class | Lack of Cohesion of Methods (LCOM2) [6] |
| | | Lack of Cohesion of Methods (LCOM3) [16] |
| Complexity | *Complexity* [7] targets the cognitive complexity of the code of a class | Weighted Method per Class (WMC) [6] |
| Coupling | *Coupling* [6] regards the inter-dependency of classes in terms of methods and attributes being used | Coupling between Objects (CBO) [6] |
| Inheritance | *Inheritance* [6], [2] encompasses parent-child relationships between classes | Depth of Inheritance Tree (DIT) [6] |
| Size | *Size* [17] measures the length or amount of source code of a class | Weight of a Class (WOC) [2] |

Table I also samples a set of six metrics assessed in the recent work mentioned above [5], which are discussed throughout this paper. All metrics are grouped (second column) according to the internal attribute they aim at capturing (first column). All six metrics become critical if their values increase after performing a change. We relied on our experiences with software development in industry and insights extracted from previous studies [14], [6] for taking this decision. In particular, we fixed the interpretation on when DIT becomes critical adopted by the authors of that recent work [5]. While they considered that DIT becomes critical as its value decreases, we consider the opposite as suggested by the paper that formalized DIT [6].

Changes affect internal attributes in three ways [5]. They can *improve an internal attribute* if the metrics used for capturing it increase or decrease towards becoming non-critical metrics. They can *keep an internal attribute unaffected* by neither increasing nor decreasing the metrics used for capturing it. They can *worsen an internal attribute* if the metrics capturing it increase or decrease towards becoming critical metrics.

### B. Refactorings

Refactoring is applying changes to improve the internal software quality [18], [19]. Developers may purely intend to enhance code structures while refactoring, or they may refactor as a means to achieve other intents, e.g. adding or enhancing features [9], [10]. Each refactoring has a type targeting the enhancement of a particular code structure. Popular refactoring types [9], [10] include Extract Method, i.e. extracting a new method from an existing one, and Move Method, i.e. moving an existing method from one class to another class. The refactoring types are implicitly associated with one or more internal attribute. This is because each refactoring type should modify the code structure and its design in such a way it changes one or more metric values.

Each internal attribute is associated with a particular subset of refactoring types [5]. For instance, the class cohesion may improve through Move Attribute and Move Method if the refactored code element (attribute or method) was at least partially the root-cause of the low cohesion. Throughout this paper, we will take some of those associations between internal attribute and refactoring types to discuss how refactorings could be used as means for managing critical attributes while evolving features.

## III. STUDY CHARACTERIZATION

### A. Problem statement

A recent study [5] quantitatively investigated the refactoring effect on five internal quality attributes. The results regarding the refactoring effect on each internal attribute were quite diverse. On the one hand, refactorings quite often enhance code structure and design, thereby helping in managing critical attributes, regardless of the applied refactoring type [5]. On the other hand, in the case of floss refactorings – which often co-occur with feature additions and enhancements – 35-55% of refactorings keep the critical attributes unaffected [5]. More critically, also in the case of floss refactorings, 9-35% of refactorings worsen these attributes.

The design of that study [5] prevented its authors from investigating the reasons behind such a high rate of refactorings that either keep internal attributes unaffected or worsen them. We hypothesize that developers only remove those degradation symptoms that matter for evolving features. Thus, developers tend to postpone or discard the removal of other, less relevant, degradation symptoms. Unfortunately, empirical evidence that supports this assumption is scarce. Previous work [20], [21] investigated the developer's perception of design smells relevance for evolving features, but we could not find similar studies in the context of critical attributes.

This paper is a first attempt to characterize how developers perceive critical attributes as relevant (or irrelevant) when evolving features. We target critical attributes associated with the five internal attributes assessed in the previous work mentioned above [5]: cohesion, complexity, coupling, inheritance, and size. The critical attributes analyzed are all the class level ones: low class cohesion, high class complexity, high class coupling, large class hierarchy depth, large class hierarchy breadth, and large class size. We discarded critical attributes at other system levels (e.g., methods) were not yet considered. We argue that developers should constantly monitor and enhance the internal quality at the class level [18], [2] and that this focus would therefore provide a valuable contribution.

### B. Research objectives

We structured our main research goal based on the Goal Question Metric goal definition template [22] as follows: *analyze* the perception of software developers on the relevance of critical attributes are relevant while evolving features; *for the purpose of* understanding; *with respect to* the individual and relative perceived relevance of the critical attributes, reasons why each critical attribute becomes relevant, and interrelations between critical attributes; *from the point of view of* developers engaged in software evolution tasks; *in the context of* two development teams evolving two systems implemented in Java within a Brazilian industry-academia R&D collaboration initiative.

We designed our industry case studies based on focus group sessions. Case studies enable observing phenomena in their natural context [11]. Focus groups, on the other hand, allow extracting experiences from the participants, promoting discussions and knowledge sharing among participants [12]. Thus, we considered that case studies with focus groups for data collection would be suitable for contextualized discussions of developers on evolving features.

### C. Context

We conducted our case study within the ExACTa PUC-Rio (https://exacta.inf.puc-rio.br) industry-academia R&D collaboration initiative, in the context of projects with Petrobras, a large company operating in the oil and gas industry. ExACTa currently has more than 70 full-time employees and delivered and evolves software-based solutions for several companies.

Our case study aims at investigating the relevance of critical attributes based on metrics computed for Java systems in a recent work [5]. As a result, we opted for selecting only teams using the Java programming language. We ended up selecting two projects as cases for our study: ship performance evaluation (Case A) and smart freight (Case B). See Section IV-B for details. Due to the subjective nature of the qualitative data analyzed and discussed throughout this work (e.g., the developer's perception of their own source code), we kept the developers associated with the development of each system anonymous. Therewith, we expect to preserve participants from any personal judgment on their perception.

Concerning the implementation, both cases followed the Lean R&D approach [23] and use Git (https://git-scm.com/systems) for performing code review via pull requests, and Azure DevOps (https://azure.microsoft.com/pt-br/services/devops/) for managing software development tasks. Both systems rely on the Spring MVC Framework (https://spring.io/) for implementing Web systems using the Model-View-Controller (MVC) architectural pattern.

## IV. CASE STUDY DESIGN

### A. Research questions

**RQ$_1$:** *What is the relevance degree of each critical attribute for evolving features from the developer's perception?*

RQ$_1$ targets two aspects of the relevance of critical attribute. First, the relevance degree of each critical attribute in isolation. In this case, we aim at understanding how important it is for developers to either mitigating or fully address a critical attribute to favor software evolution. Second, the relative relevance of the six critical attributes altogether: low class cohesion, high class complexity, high class coupling, large class hierarchy depth, large class hierarchy breadth, and large class size. Thus, we expect to understand what critical attributes have the highest priority when it comes to evolving features.

**RQ$_2$:** *What are the reasons behind considering a critical attribute as relevant for evolving features?*

RQ$_2$ explores the circumstances that make certain critical attributes actually relevant for evolving features. We ask participants of each focus session group to argue why each critical attribute deserves special attention while either adding or enhancing features. This knowledge is useful for supporting decision-making in development teams with little time for delivering features. Indeed, there might be hundreds of stakeholders' demands for developers to worry about while performing software evolution [24]. Thus, they may find certain critical attributes worth managing in the detriment of other critical attributes.

### B. Case and subject selection

**Case A:** This case consists of implementing a software solution for shipping logistics management purposes at Petrobras. Such management occurs through the integration of multiple systems, which allow practitioners to identify and cope with situations in which a chartered ship is not performing as expected or is not available for delivering a service. The proposed system heavily depends on computing business rules, e.g. with respect to payments and available fuel computation. Four employees of Petrobras often participate with feedback and advice during the software development process. They support the R&D team in Scrum planning and review cycles whenever possible.

**Case B:** This case consists of implementing a software solution for handling freight calculation. It is responsible for assisting practitioners at Petbrobras in predicting freight prices to transport materials via road transport. The system relies on a large data set, including information on truck size and distances among cities. The proposed system also estimates fair prices to transport materials between refineries and so forth. Thus, this system is also heavily grounded in business rule computation, data processing, and model-based prediction.

Table II provides an overview of the participant background for each case, which we collected from an online Characterization Form sent to participants minutes before the start of each focus group session. Regarding the highest education degree, while Case A has a PhD, both cases counted on the participation of one Master and one Specialist in knowledge areas related to Computer Science. Although participants of Case A have 4.33 years of experience on average against 8.33 years for Case B, participants in both cases have participated in the development of 5 systems on average.

Table II
PARTICIPANT BACKGROUND COLLECTED VIA BACKGROUND FORM

| Question (Q) | Case A | | | Case B | | |
| --- | --- | --- | --- | --- | --- | --- |
| | A1 | A2 | A3 | B1 | B2 | B3 |
| Q1: Highest education degree | MSc | Specialist | PhD | Specialist | MSc | BS |
| Q2: Years of industry experience | 5 | 2 | 6 | 5 | 10 | 10 |
| Q3: Number of software projects | 7 | 4 | 4 | 2 | 6 | 7 |
| Q4: Familiarity with software metrics | Q4.c | Q4.c | Q4.b | Q4.d | Q4.c | Q4.d |
| Q5: Concerned with improving quality | Agree | Agree | Indifferent | Strongly agree | Agree | Strongly agree |
| Q6: Familiarity with internal attributes | Q6.e | Q6.c | Q6.b | Q6.d | Q6.d | Q6.d |

Q4.b: I've heard about them but I am not so sure what they are
Q4.c: I've a general understanding, but do not use them in my sw. projects
Q4.d: I've a good understanding, and use them in my sw. projects sometimes
Q6.b: I've heard about them but I am not so sure what they are
Q6.c: I've a general understanding, but do not analyze them in my sw. projects
Q6.d: I've a good understanding, and analyze them in my sw. projects sometimes
Q6.e: I've a strong understanding, and analyze them in my sw. projects frequently

The Characterization Form asked participants to report on their familiarity with software metrics (Q4). According to the data of Table II, two participants of Case A said they have heard of metrics but are not really sure about what metrics mean (Q4.b), while one participant said to have a general understanding but does not use metrics in his projects (Q4.c). Case B participants claimed to be slightly more familiar with metrics: while one participant said to have a general understanding but not to use metrics (Q4.c), two participants said they have a good understanding and use metrics sometimes (Q4.d). We also asked participants on how much they are concerned with improving the quality of source code in their systems (Q6). While participants of Case A simply agreed with this statement, participants of Case B showed to be significantly more concerned about this matter. Judging by the lower familiarity of participants in Case A with metrics, this result is reasonable.

Finally, the Characterization Form asked about the familiarity of participants with the concept of internal attributes. As data of Table II suggests, participants of Case A oscillated a lot in terms of expertise with internal attributes: one of them only heard about it (Q6.b), another one said to have a general understanding but not analyzing internal attributes in his projects (Q6.c), and the last one said to have a strong understanding and analyzing internal attributes frequently (Q6.e). One could say these participants know at least a little about internal attributes because of the metrics they are familiar with. Conversely, all participants of Case B informed to have a good understanding and analyze internal attributes occasionally (Q6.d).

Participants of Case A are less familiar with internal quality and its management than participants of Case B, but all participants are considerably experienced in software development.

### C. Data collection procedures

Figure 1 depicts the procedures adopted for collecting data throughout the case study. We organized these procedures in three major phases, which we describe below.

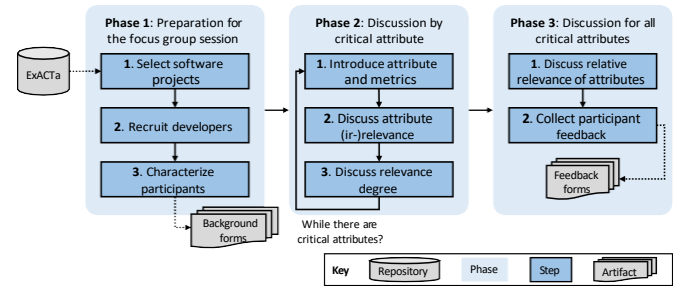

Figure 1. Data Collection Procedures

**Phase 1:** *Preparation for the focus group session* – Besides defining what cases would be assessed and recruiting participants to engage in discussions, we collected the background characterization information. This phase has the three procedures below.

**Procedure 1.1:** *Select software projects* – We contacted a development project manager at the R&D initiative asking for software projects whose systems are implemented using mainly Java. We made this decision because a recent work [5] focused on Java and to assure that participants are minimally familiar with object-oriented languages, which permeate all this work (e.g., the refactoring types and metrics explored here are mostly applicable to these languages). We ended up selecting two projects (Case A and Case B).

**Procedure 1.2:** *Recruit developers* – For each project, we asked for permission to invite developers for participating in our study. Due to the intensively collaborative nature of the R&D projects, both cases share developers, which contribute in the development of multiple systems. We opted for selecting two independent sets of participants, one per system. No participant engaged in two focus group sessions, even though they may have contributed to the development of both systems. The project manager played an essential role in defining what developers are more active in the development of each system. We recruited three participants per case.

**Procedure 1.3:** *Characterize participants* – We carefully designed and revised our background Characterization Form

aimed at collecting basic information on the participants' expertise. Our major goal was profiling each case so we could better interpret our study results. We opted for a short and simple form in order to prevent participants from being tired or discouraged to participate in discussions right after filling the form. As shown in Table II, we collected data on the participant education (Q1), experience with software development in industry (Q2 and Q3), familiarity with two key concepts of this work, i.e. software metrics (Q4) and internal attributes (Q6), and the concern of developers in improving code quality (Q5).

Before discussing **Phase 2** and **Phase 3**, we explain the online environment used for promoting discussions on critical attributes. Figure 2 depicts the virtual template that we carefully designed using the MURAL online tool[1]. The MURAL team kindly granted us with a free workspace at the MURAL for Education program. Each session started with one empty version of our designed virtual discussion template to be shared by all participants of the same session. This template has seven well-defined sections. Sections A to F aimed at driving the discussion regarding each of the six critical attributes. Section G aimed at driving the discussion on the relative relevance of all six critical attributes.

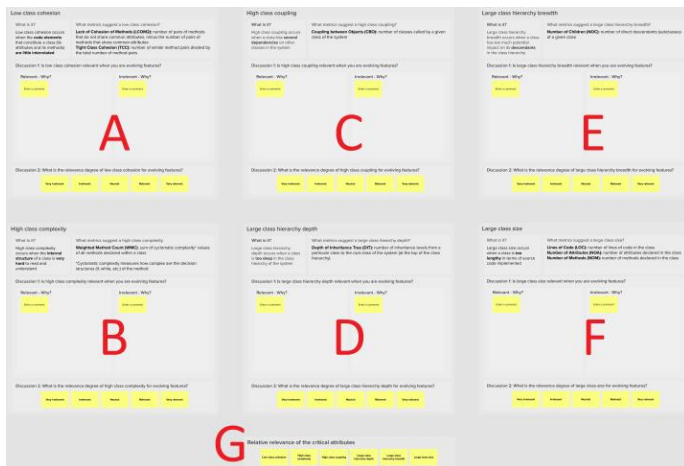

Figure 2. Template of Focus Group Session Defined at MURAL

Figure 3 depicts only Section A designed for discussing low class cohesion. Section A1 contains a short description of the critical attribute based on the literature [5], [18], [2]. Section A2 provides the participants with examples of metrics aimed at capturing the respective critical attribute. Section A3 was designed for developers to add notes on why the critical attribute is relevant for evolving features. Section A4 is similar but focused on why critical attributes may be irrelevant for evolving features. Finally, Section A5 is designed for capturing the relevance degree of the critical attribute based on a five-point scale: very irrelevant, irrelevant, neutral, relevant, and very relevant.

**Phase 2:** *Discussion per critical attribute* – This is the first phase associated with focus group session itself. We collected all data regarding the developer's perception of critical attributes as relevant (or irrelevant) for evolving features. However, in this phase, each critical attribute is discussed in isolation. We defined the three procedures below.

[1] https://www.mural.co/

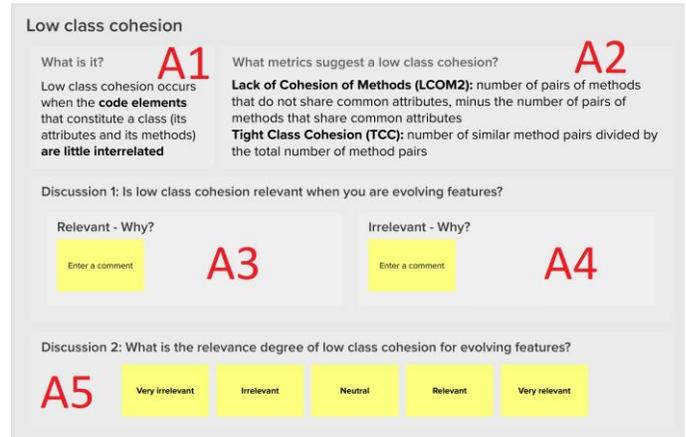

Figure 3. Section Dedicated to Discussing Low Class Cohesion

**Procedure 2.1:** *Introduce attribute and metrics* – The discussion on each critical attribute starts by providing the participants with a short definition of the critical attribute based on the literature [5], [18], [2]. After that, we provided them with a few examples of metrics designed for capturing the respective internal attribute. We sampled metrics arbitrarily based on our notion of metrics that developers could understand more easily. For instance, for exemplifying Lack of Cohesion we opted for (LCOM2) [6] rather than LCOM3 [16] because the latter implies explaining concepts like disjoint components in a graph.

**Procedure 2.2:** *Discuss attribute (ir-)relevance* – We asked the participants to elicit reasons why each critical attribute is relevant (or irrelevant) for evolving features. Each reason should be documented as a note in the appropriate section: Section A3 for relevant and Section A4 for irrelevant. From time to time, we reminded participants that evolving features include adding new features as much as enhancing existing features of the system. In addition, we constantly recommended participants to share knowledge and experiences surrounding each critical attribute, especially when discussions lost intensity. All participants were asked to collaborate with circumstances where mitigating or fully addressing a critical attribute is important for facilitating software evolution.

**Procedure 2.3:** *Discuss relevance degree* – After discussing why a critical attribute is relevant (or irrelevant) for evolving features, the facilitator of the focus session group promoted a recap. Each note on the (ir-)relevance of the critical attribute was read out loud. Whenever the facilitator felt that a note is poorly written, he asked the participants to provide further considerations on the note. At the end of this procedure, we asked each participant to assign one vote to the relevance degree of the critical attribute using a five-point Likert scale.

**Phase 3:** *Discussion for all critical attributes* – After discussing each critical attribute in isolation, the focus session group ended with a discussion about the relative relevance of critical attributes. The procedure of this phase follows.

**Procedure 3.1:** *Discuss relative relevance of attributes* – We asked participants to rank those critical attributes that matter the most while evolving features. Each participant received five votes. These votes were meant to be distributed throughout the six critical attributes. We arbitrarily chose to assign five votes per participant in order to prevent them from assigning one vote for each critical attribute, thereby making it hard to conclude anything on the relative relevance.

Each focus group session was conducted online via a Zoom Meeting. We kept video and audio records of both sessions to support the analysis of data provided by the participants. We often accessed the video and audio records for understanding what developers meant with each note. The focus group sessions for Case A and Case B each lasted approximately two and a half hours.

### D. Data analysis procedures

We asked participants of each focus group session to provide us with reasons why each critical attribute is relevant (or irrelevant) for evolving features. We collected these reasons through notes posted by the participants in the session's virtual mural. Aimed at analyzing these reasons, we first transcribed all notes exactly as they were written by the participants into a spreadsheet. Based on our impressions as facilitators of the focus group sessions, and after watching the video and audio records, we rewrote the notes aimed at fixing typos, filling communication gaps (e.g. omitted words), and making the target critical attributes of each note explicit. Finally, we translated the notes from Brazilian Portuguese to English.

We applied thematic synthesis [25] on the qualitative data for Case A and Case B) separately. First, we separated the notes regarding the relevance of all critical attributes from those regarding irrelevance. Second, for each set of notes (relevant and irrelevant), we grouped them according to their core theme, i.e. fine-grained themes discussed throughout each note. Third, we grouped these core themes into macro-themes, i.e. themes that are more comprehensive than the core themes. Fourth, we separated those macro-themes into two categories: the ones regarding *Code Structure and Design* and the ones regarding *System Functionality*.

## V. RELEVANCE OF CRITICAL ATTRIBUTES FOR EVOLVING FEATURES (RQ$_1$)

### A. Relevance of critical attributes per case

Figure 4 depicts how many participants voted for a certain degree of relevance with respect to each critical attribute under investigation. We grouped the results by case: Case A data in the left and Case B in the right.

Regarding Case A, three critical attributes are ultimately perceived as relevant by the developers while evolving features: low class cohesion, high class complexity, and large class size. These are the only critical attributes for which no participant reported perceptions as either neutral or (very) irrelevant. Data of Table II suggests that participants of Case A are less familiar with metrics and internal attributes. Thus, one could speculate that low class cohesion, high class complexity, and large class size are intuitive degradation symptoms – and potentially harmful to evolving features.

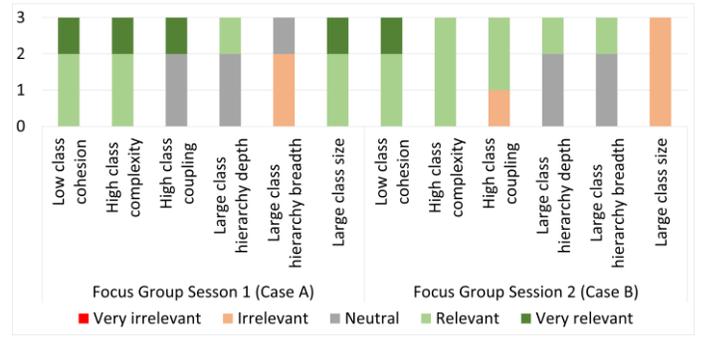

Figure 4. Relevance of Critical Attributes per Case

With respect to Case B, our study results changed considerably. First, only low class cohesion and large class complexity are reportedly relevant for evolving features. Curiously, in the opposite way of Case A, large class size was ultimately considered irrelevant while performing software evolution. None of the participants assigned the very irrelevant degree, but they also did not assign any relevant degree. This may be due to Case B participants being the most familiar with metrics and internal attributes (Table II). Maybe they are experienced enough to acknowledge that certain large classes are acceptable depending on factors such as the system domain and the inherent difficulty of a business rule.

### B. Relevance of critical attributes for both cases

Figure 5 depicts the overall developer's perception on how relevant each critical attribute is, regardless of the case.

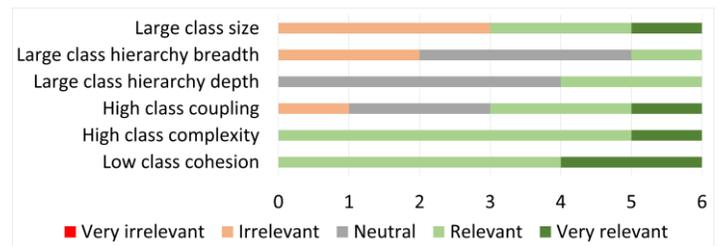

Figure 5. Relevance of Critical Attributes for Both Cases

Two critical attributes are ultimately relevant for developers: low class cohesion and high class complexity. The other four critical attributes are not necessarily relevant while evolving features: high class coupling, large class hierarchy depth, large class hierarchy breadth, and large size. Curiously, the aggregated data shows that developers are not exactly sure whether large class hierarchy depth. The high rate of neutral votes suggest that this critical attribute is not even an issue when debating and performing software evolution.

## VI. Reasons behind the (ir-)relevance of critical attributes (RQ2)

### A. Overall results for Case A

Table III lists the notes provided by participants on why each critical attribute (first column) is either relevant or irrelevant for evolving features. The second column distinguishes notes about the attribute relevance or irrelevance.

Table III
NOTES ON (IR-)RELEVANCE OF CRITICAL ATTRIBUTES FOR CASE A

| Critical Attribute | Total of Notes | Notes specifically related to... | |
|---|---|---|---|
| | | Relevance | Irrelevance |
| High class complexity | 6 | 5 | 1 |
| High class coupling | 6 | 3 | 3 |
| Large class hierarchy breadth | 6 | 2 | 4 |
| Large class hierarchy depth | 7 | 4 | 3 |
| Large class size | 10 | 6 | 4 |
| Low class cohesion | 9 | 7 | 2 |

Large class size was the most discussed critical attribute in terms of number of notes, which is not surprising because it is a top-three most relevant attribute according to Figure 4. This is interesting the usefulness of size metrics for assessing the internal software quality is quite debatable [26], [27], [2], [16]. On the one hand, participants reported, for instance that "large class size makes it hard to maintain source code" and "large class size increases error proneness of the source code," both topics discussed by previous work [26], [27]. On the other hand, participants mentioned that "large class size is irrelevant when developers deal with urgency in program delivery."

The second most discussed critical attribute is low class cohesion, which is also a top-three most relevant attribute according to Figure 4. This is an attribute whose applicability in measuring internal quality has been shown in different development scenarios [28], [6]. Participants said that "high class cohesion facilitates source code reuse," something that previous work has assessed [28]. Participants also said that "low class cohesion makes it hard to find errors," which is a recurring argument through the responses of both cases (Case A and Case B). On the other hand, participants discussed that "low class cohesion is irrelevant when the time to delivering code is short," which has been discussed with respect to large class size as well. Curiously, in the particular case of low class cohesion, participants seem to have more arguments on the relevance of this critical attributes for evolving features.

The remaining critical attributes were also significantly discussed. Large class hierarchy depth had many notes regarding its relevance for evolving features, while high class complexity, high class coupling, and large class hierarchy breadth. Curiously, high class complexity had the highest number of notes on its relevant while evolving features; this results is interesting because high class complexity is the last top-three most critical attribute according to Figure 4.

Regarding the two critical attributes associated with inheritance, participants tended discuss irrelevance by means of the practical usefulness of class hierarchies in a system. Examples of quotes that illustrate this issue are "large depth is irrelevant when it allows reusing code located at the highest hierarchical levels" and "large breadth is irrelevant when reusing properties used by all entities of the Entity Relationship Diagram." Quotes like these justify, at least in parts, why the majority of participants assigned a neutral or irrelevant degree for the relevance of both attributes (cf. Figure 4).

During each focus group session, we constantly stimulated participants to report as many aspects of either relevance or irrelevance by critical attribute. Judging by the considerable number of arguments favor and against the relevant of all attributes, we concluded that our effort in promoting a healthy and productive discussion among participants paid off.

### B. Thematic synthesis results for Case A

Figure 6 depicts our results for the thematic synthesis procedures applied to notes on why critical attributes are *relevant* for evolving features. The root note of the tree corresponds to the major theme, i.e. the relevance of all critical attributes altogether. The first intermediate level includes the two categories mentioned above (Code Structure and Design and System Functionality). The second intermediate level includes the macro-themes. We assigned in brackets the critical attributes associated with each micro-theme or macro-theme whenever the node corresponds to a leaf from the tree – i.e. the node has no variants. We have found seven macro-themes, four associated with code structure and design. The leaves correspond to the nine micro-themes.

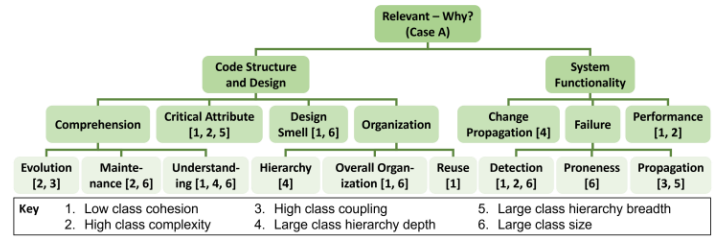

Figure 6. Themes on Why Attributes are Relevant for Case A

Regarding Code Structure and Design, participants mentioned that critical attributes are relevant for: *Comprehension*, i.e., the ability of reading and understanding the code elements; *Critical Attribute*, i.e. analyzing and reasoning about other critical attributes; *Design Smell* i.e. assessing or reasoning about the occurrence of Fowler-like design smells [18]; and *Organization*, i.e. the way how code elements are organized within the source code structure. The micro-themes have intuitive names, but note *Overall Organization* includes aspects neither associated with *Hierarchy* nor with *Reuse*.

On System Functionality, participants said that critical attributes are relevant for: *Change Propagation*, i.e. critical attributes may spot cases in which certain changes are unexpectedly or undesirably propagated throughout the system; *Failure*, i.e. the occurrence of bugs, faults, or failures; and *Performance*, i.e. aspects of the system performance such as the speed to respond to requests.

Complementarily, Figure 7 is a tree of themes on why critical attributes are *irrelevant* while evolving features (cf. tree root). The first intermediate level corresponds to the categories Code Structure and Design and System Functionality. The second intermediate level corresponds to seven macro-themes identified, four of them associated with system functionality. The leaves are the five micro-themes identified. We assigned in brackets the critical attributes associated with each micro-theme or macro-theme whenever the node corresponds to a leaf from the tree – i.e. the node has no variants.

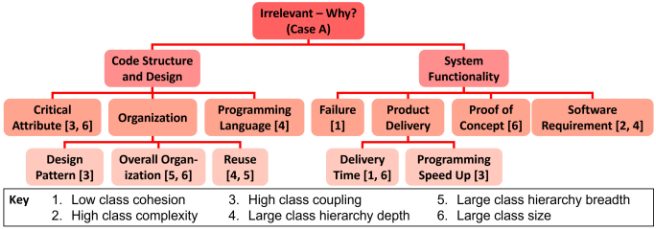

Figure 7. Themes on Why Attributes are Irrelevant for Case A

With respect to Code Structure and Design, participants mentioned that critical attributes, in general, are irrelevant for: *Critical Attribute*, i.e. analyzing and reasoning about other critical attributes; *Organization*, i.e. the way how code elements are organized within the source code structure; and *Programming Language*, i.e. aspects derived from the syntax, structure, and features provided by the programming used during software evolution. The micro-themes have quite intuitive names, but we highlight that *Design Pattern* refers to Gamma-like design patterns [29]. The participants argued that adopting certain patterns might lead to critical attributes. Despite their theoretical or practical impact on the internal software quality, the critical attributes are either irrelevant or cannot be managed.

Regarding System Functionality, participants reported that critical attributes are irrelevant for: *Failure*, i.e., certain circumstances associated with bug fixing – in this case, when "fixing bugs in legacy code"; *Product Delivery*, i.e., aspects of delivering a system; *Proof of Concept*, i.e., aspects associated with the implementation of source code particularly aimed at proving concepts to stakeholders during the iterative development cycles of agile processes; and *Software Requirements*, i.e., requirements in general.

Via the **Critical Attribute** macro-theme, we have found interesting insights on the interrelation between different critical attributes. About *relevance*, participants of Case A said that: i) "high class complexity leads to high class coupling," ii) "large breadth may increase the class complexity," and iii) "non-cohesive classes tend to be larger than necessary." In summary, we found three tuples of perceived interrelations: i) (high class complexity, high class coupling), ii) (large class hierarchy breadth, high class complexity), and iii) (low class cohesion, large class size), respectively. This result could support the design of refactoring tools that, aimed at enhancing code structures, optimize certain critical attributes that are interrelated.

Regarding *irrelevance*, participants of Case A said that: i) "high class coupling is acceptable when coding a highly coupled entity of the Entity Relationship Diagram" and ii) "large class size is irrelevant when the methods are naturally very complex." Thus, we found two tuples of interrelations: i) (high class coupling, high entity coupling) associating attributes at the levels of class and Entity Relationship (ER) model, and ii) (large class size, high method complexity) associating attributes at the levels of class and method, respectively. Again, these results could drive the design of novel refactoring tools for enhancing code structure and design.

**A parallel between the relevance of critical attributes and design smells:** Earlier in Section IV-A, we discussed that critical attributes may help in detecting design smells [18], [2]. Curiously, the results of Figures 6 suggest that critical attribute are closely associated with certain design smells. Especially, Case A participants suggest that low class cohesion and large class size are relevant because of their association with Duplicated Code, i.e. different code snippets realizing the same feature, and Large Class, i.e. a class overloaded with several features. This result is interesting because a previous study [20] suggests that developers often perceive Complex Class, Large Class -equivalent to God Class [1], Long Method, and Spaghetti Code (all associated with low class cohesion and large class size) as potentially harmful to software evolution.

### C. Overall results for Case B

Table IV lists the number of notes on why each critical attribute is either relevant or irrelevant for evolving features.

Table IV
NOTES ON (IR-)RELEVANCE OF CRITICAL ATTRIBUTES FOR CASE B

| Critical Attribute | Total of Notes | Notes specifically related to... | |
| --- | --- | --- | --- |
| | | Relevance | Irrelevance |
| High class complexity | 4 | 3 | 1 |
| High class coupling | 7 | 4 | 3 |
| Large class hierarchy breadth | 3 | 2 | 1 |
| Large class hierarchy depth | 5 | 3 | 2 |
| Large class size | 7 | 2 | 5 |
| Low class cohesion | 6 | 3 | 3 |

Large class size tied with high class coupling as the critical attributes with the highest number of notes. Curiously, these critical attributes were the only ones to have at least one vote for irrelevant (Figure 4). As previously discussed with respect to Case A, the usefulness of size metrics for assessing the internal software quality has been debated by previous work with mixed opinions [26], [27], [2], [16]. This debate is reflected by the highest number of votes for irrelevant from the entire case study (Figure 5). On the one hand, participants said that "large size is relevant if the programming screen size is small" and because "large size rarely occurs in isolation" in terms of problems associated with internal software quality. On the other hand, participants also said that "large size is irrelevant in utility classes" and "large size is irrelevant if the integrated development environment (IDE) can collapse large source code blocks." These comments particularly suggest that large class size a problem of the development environment rather than a system problem.

Regarding high class coupling, participants also showed different perspectives on relevance and irrelevance. On the one hand, participants reported that "high class coupling is relevant when implementing fault tolerance/error handling." They also were very specific on the metrics used for computing this critical attribute, with statements like "high class coupling is relevant when CBO is high but the class is coupled with classes at different program levels." CBO is the Coupling between Objects metric of Table I, which we displayed in our virtual mural during the focus session group for exemplification. By the way, with "program level" the participants clarified that they refer to a system package or module. On the other hand, participants said that "high class coupling [is irrelevant because it] may support source code reuse" and, in opposition to the previous case, "high class coupling is irrelevant when CBO is high but the class is coupled with classes at the same program level." It is worth mentioning that, although is has mostly been seen as relevant for evolving features (Figure 4), this critical attribute received one vote for irrelevant.

The third most discussed critical attributes is low class cohesion. Participants of Case B reported that "low class cohesion makes it hard to maintain a program" and "low class cohesion makes it hard to track errors." On the other and, from the developer's perception, "low class cohesion is irrelevant if it affects an utility class" and "low class cohesion is irrelevant in very small programs," for instance. Comments like these suggest that the relevance of low class cohesion strongly depends on what the system implements. If the system is too simple or the class provides features to the whole system, low class cohesion is acceptable. Regardless of that, this critical attribute is curiously the one with most votes for relevant (with one vote for very relevant) – according to data of Figure 5.

The other three critical attributes – high class complexity, large class hierarchy depth, and large class hierarchy breadth – were less discussed in comparison with the same attributes in Case A. Curiously, the overall perception of developers in Case B are quite different and valuable. Regarding the attribute relevance, participants report that "high class complexity makes it hard to implement new business rules," "large depth makes it hard to know where to implement a new program feature," and "large breadth is relevant if child classes redefine the concrete behavior inherited from their parent class." This notes add up as they confirm how different critical attributes may hinder feature additions and enhancements, which are the basis of software evolution. In addition, all the three critical attributes received at least one vote for relevant (Figure 4). It is worth mentioning that the two critical attributes regarding inheritance received neutral votes. This results suggest that, for developers considerably concerned on internal software quality (Table II), inheritance is not a major concern.

*D. Thematic synthesis results for Case B*

Figure 8 is a tree showing results for the thematic synthesis procedures applied to notes on why critical attributes are *relevant* during software evolution. The root note of the tree corresponds to the major theme, that is, the irrelevance of all critical attributes altogether. The first intermediate level corresponds to the two major categories of themes: Code Structure and Design and System Functionality. The second intermediate level corresponds to the macro-themes. We have derived seven macro-themes where the majority (five of them) is associated with code structure and its design. The leaves are the seven micro-themes found in total. We assigned in brackets the critical attributes associated with each micro-theme or macro-theme whenever the node corresponds to a leaf from the tree – i.e. the node has no variants.

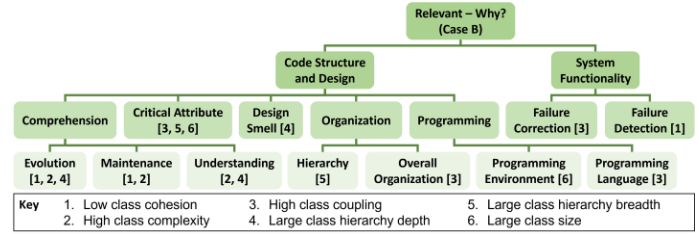

Figure 8. Themes on Why Attributes are Relevant for Case B

With respect to Code Structure and Design, participants mentioned that critical attributes in general are relevant for: *Comprehension*, i.e., the ability of reading and understanding the code elements; *Critical Attribute*, i.e. analyzing and reasoning about other critical attributes; *Design Smell* i.e. assessing or reasoning about the occurrence of Fowler-like design smells [18]; *Organization*, i.e. the way how code elements are organized within the source code structure; and *Programming*, i.e. aspects of programming a system that may affect how the internal software quality is perceived. All micro-themes received intuitive names with no need for further explanation.

Regarding System Functionality, participants reported only two macro-themes; *Failure Correction*, i.e. the ability to fix bugs, faults, or failures affecting the system behavior; and *Failure Detection*, i.e. the task of tracking unexpected system behaviors realized by bugs, faults, or failures in a system.

Figure 9 depicts the themes derived the topic of why critical attributes are *irrelevant* while evolving features (see the tree root). The first intermediate level corresponds to the categories Code Structure and Design and System Functionality. The second intermediate level corresponds to five macro-themes identified, three of them associated with code structure and its design. The leaves correspond to four micro-themes. We assigned in brackets the critical attributes associated with each micro-theme or macro-theme whenever the node corresponds to a leaf from the tree – i.e. the node has no variants.

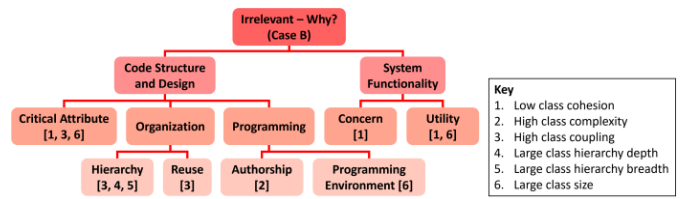

Figure 9. Themes on Why Attributes are Irrelevant for Case B

With respect to Code Structure and Design, participants reported that critical attributes in general are relevant for: *Critical Attribute*, i.e. analyzing and reasoning about other critical attributes; *Organization*, i.e. the way how code elements are organized within the source code structure; and *Programming*, i.e. aspects associated with activity of programming a system. All micro-themes have intuitive names. Still, it is worth mentioning that *Authorship* refers to the authorship of source code implemented by the developers.

Regarding System Functionality, participants reported that critical attributes are irrelevant in cases associated with: *Concern*, i.e. nature of the features realized by the system; and *Utility*, i.e. classes of the system that serve as feature providers to the majority of the system – also known as utility classes.

Analyzing the **Critical Attribute** macro-theme, we derived interesting insights on the interrelation between different critical attributes. About *relevance*, participants of Case B said that: i) "high class coupling is relevant when affecting non-cohesive classes," ii) "large depth makes large breadth worse," and iii) "large size rarely occurs in isolation." Based on these quotes, we identified three tuples of interrelations: i) (high class coupling, low class cohesion), ii) (large class hierarchy depth, large class hierarchy breadth), and iii) (large class size, any critical attribute), respectively. We did not find any similarities with the interrelations of Case A and Case B.

Regarding *irrelevance*, participants of Case B said that: i) "high class coupling is irrelevant when the class is highly cohesive," ii) "large size is irrelevant if the affected class is not complex," iii) "large size is irrelevant if the methods are cohesive," iv) "large size rarely occurs in isolation," and v) "low class cohesion is irrelevant in very small programs." Thus, we found five tuples of interrelations: i) (high class coupling, low class cohesion), ii) (large class size, high class complexity), iii) (large class size, low method cohesion), iv) (large class size, any critical attribute), and v) (low class cohesion, high system size), respectively. We could not find any similarities with the interrelations of Case A and Case B.

**A parallel between the relevance of critical attributes and design smells:** The results of Figure 6 also suggest that critical attribute are closely associated with some design smells. As a complement to the data of Case A (Section VI-B) Case B participants suggest that large class hierarchy depth is relevant because it may lead to Duplicated Code. This result stands out because a previous study [21] suggests that Duplicated Code is often perceived as harmful to software evolution. Our study results confirm, to some extent, the practical relevance of at least five traditional design smells: Complex Class, Large Class/God Class, Long Method, and Spaghetti Code for Case A and Duplicated Code for Case B.

## VII. RELATED WORK

Some studies have investigated whether developers perceive anomalous metric values as useful degradation symptoms [30], [27], [31], [32]. One study [31] assessed factors leading to a high attractiveness of open source systems to new developers. Results suggest that developers may be discouraged to contributing to systems with anomalous values of complexity metrics. Other studies [30], [32] captured the developer's perception on the usefulness of coupling metrics. Results suggest coupling metrics solely based on the analysis of code structures, such as CBO, are less effective than those derived from semantic aspects of the system (e.g., feature location). Another study [27] concludes that developers often perceive anomalous metric values for complexity and size as indicators of unclear code (i.e., code that is potentially hard to understand and modify).

Each metric aims at capturing a particular internal attribute (Section II-A). Thus, one could assume the aforementioned study results provide hints on how relevant the critical attributes are for software evolution in industry. Still, we could not find industry-focused studies on the perception of multiple class-level critical attributes as relevant while evolving features. We only found studies on the developer perception of design smells as useful degradation symptoms [20], [21].

A few insights on critical attributes appeared in the smell-centered studies because certain design smells are defined by combining multiple critical attributes [18], [2]. Particularly, these studies suggest that Large Class and Long Method are design smells that make it hard for developers to understand and modify code [20], [21]. This result stands out because both design smell types are often detected by combining low cohesion, high complexity, and large size [33], [8], which were perceived by our study participants as relevant (Figure 5).

## VIII. THREATS TO VALIDITY

**Construct Validity:** We defined the case study protocol and artifacts based on empirical software engineering research guidelines, e.g., [25] and [11]. We used an extensive recent study [5] as a reference for defining our study steps and procedures. Thus, we expected to support the proper data collection and analysis (Section IV-C). All six critical attributes investigated rely on the five internal attributes assessed in an extensive study [5]. By focusing on the same set of attributes, we expected to support the comparison of both quantitative and qualitative studies. Two researchers contributed with insights on how to organize and conduct the focus group sessions. Finally, our study targeted six critical attributes at the class level. These attributes have been typically used for monitoring the internal software quality [5]. We are aware that the participant background may have affected our study results, although we conducted a careful recruitment process (see Section IV-B).

**Internal Validity:** Aimed at stimulating participants to engage in the focus group sessions, we informed our agreement in donating food to charity for each engaged participant. Each focus group session lasted no longer than two hours and a half. During the focus groups sessions, we instructed participants on the formal definition of each critical attribute and the metrics frequently used for capturing them. Therewith we expected to normalize the participants' knowledge in preparation for a fruitful discussion. We are aware that our interactions with the study participants may have biased their reported perceptions, although we did our best to avoid sharing our thoughts and perceptions during the sessions. Finally, we kept video and audio records of each focus group session, with the permission of all participants, to support our posterior data collection and analysis, helping to avoid missing and incorrect data.

**External Validity:** We opted for performing case studies, which typically encompass only a small set of subjects and cases [11]. One could argue that such a limited set would have hindered the derivation of relevant study results on developers' perceptions of critical attributes. Although we partially agree with this argument, focus groups are not intended for large-scale analyses [12]. Rather, these studies aim at promoting discussions among a few participants in such a way that controlling their participation and reaching qualitative depth (e.g., explanations) becomes possible. Finally, the two systems analyzed in this work are mainly implemented in the Java programming language, which is largely used in industry.

**Reliability:** One could argue that the developer's perception may not be the best way of establishing interrelations of critical attributes. Particularly, there may be other interrelations not covered by our study or neglected by the developers for some reason. Still, we believe that considering developers' perceptions makes sense for different reasons. For instance, developers argue that low class cohesion and high class coupling are interrelated, which is reflected in recurring strategies for detecting design smells such as Large Class [18], [2]. Similar reasoning applies to other interrelations, such as high class complexity and large class size [18], [2]. Interrelations reported by the participants may not saturate all possibilities but, still, their perceptions provide meaningful insights into how they can be best assisted aiming at managing critical attributes in practice. To further improve the reliability and allow the replication of our analyses, all artifacts and data, including the complete transcriptions of the focus group sessions, are available in our online material [13].

IX. STUDY IMPLICATIONS

**Implication to Practitioners:** *Existing techniques could help in mitigating or fully addressing critical attributes while evolving features* – The focus group sessions aimed at confirming and complementing preliminary insights of previous studies [20], [21] on the importance of addressing critical attributes for facilitating software evolution. Our results include the validation of low class cohesion and high class complexity as ultimately relevant from the developer's perspective. Existing techniques for detecting design smells – which are usually combinations of two or more critical attributes – could be useful for assisting developers in analyzing critical attributes in their systems. There is a myriad of options in the literature for this particular purpose [1], [3], [19], [34].

In addition, the literature of anomalous metric values and critical attributes is diverse and comprehensive [14], [15], [5], [2]. Still, our Background Form in particular (Section IV-B) reminded us that developers may not be sufficiently aware of the techniques already proposed for supporting the analysis of critical attributes in practical settings. Raising such awareness is fundamental for developers to discuss and manage critical attributes, especially in cases of short time for delivery products – a recurring issue (cf. Table III and Table IV).

**Implication to Researchers:** *Recommender systems should incorporate mechanisms for driving changes also considering practical developer intents, rather than the pure enhancement of code structures* – Recommender systems, e.g. [3], [34], assist developers in enhancing code structures [34] and evolving software architectures [3]. In both cases, the traditional refactorings of Fowler's Refactoring book [18] are employed towards a facilitated software development in general. These tools rely on the assumption that code structures and design that are free of degradation symptoms are favorable to adding or enhancing features. This assumption is not incorrect *per se*. However, the ideal code structure and design suggested by these tools may be either unnecessary or too costly to achieve.

Our study results suggest that, during software evolution, developers tend to concentrate effort on applying only those changes necessary to achieve their major intent. While existing recommender systems typically suggests at once dozens of changes, their practical adoption sounds unrealistic at times. This observation is especially valid when the developer wants to add or enhance features very locally in the code. Our results could help researchers in re-designing tools for addressing degradation symptoms – including critical attributes with fewer changes and focused on problems that actually affect the tasks of evolving features. Tool designers could incorporate our insights on the practical relevance of critical attributes, as their interrelations, to assist disciplined refactorings aimed at mitigating or fully addressing critical attributes.

X. FINAL REMARKS

We conducted a qualitative case study on the perception of developers from two different development teams on the relevance of critical internal quality attributes when evolving software features. We also elicited and conducted a thematic analysis of reasons why developers find each critical attribute relevant. We reveal some interrelations of critical attributes as perceived by the developers. In particular, we found that low class cohesion and high class complexity are perceived as highly relevant and that they are interrelated with other critical attributes. Refactoring tools such as [3], [34] optimize multiple critical attributes at once, while we found that not all critical attributes are equally relevant to developers. We encourage considering the perceived relevance of critical attributes while designing tools as a means to better meet the developers' expectations.

While the thematic analysis of the two selected cases provided valuable insights into the perceived relevance of critical internal quality attributes when evolving software features, there are inherent limitations of this first qualitative industry case study on the topic and we plan to replicate our study. In particular, we aim at higher diversity and representativeness [35], e.g., by also recruiting participants working on projects implemented in other programming languages.